# Process Description

# of

# COM Object Life Cycle

by

**Emil Vassev**

A Technical Article
March 2002

# ABSTRACT

The objective of this article is to provide for the reader a basic description of all the steps involved in the COM object life-cycle process. COM is a software technology and process performer. The first section briefly introduces the Component Object Model (COM), considering the process of the COM object life cycle as the baseline of all COM issues. The second part describes in detail the basic steps of the process – client request, server location, object creation, interaction, and disconnection. A brief description is given for the components involved in each step. Finally, the third section provides a brief conclusion summarizing all the process steps.

# 1. GENERAL INTRODUCTION

The COM object life-cycle process is a process of resource maintenance and component interaction performed by Component Object Model (COM). COM is a general software architecture developed by Microsoft that provides a framework for integrating software components. This framework allows developers to build systems by assembling reusable components. By defining an application-programming interface (API), COM allows the creation and integration of components in custom applications or allows diverse components to interact [1]. The interaction is possible only if the components "adhere to a binary structure specified by Microsoft" [1]. Components written in different programming languages can interoperate if they adhere to this binary structure.

The process of the COM object life-cycle is the most common process performed by COM, and it takes place in any COM issue. While earlier Microsoft Operating Systems, as Windows 95-98 and Windows NT, were more or less COM independent, COM has been fully integrated in Windows 2000. COM is also the base for such technologies like OLE (Object Linking and Embedding), ActiveX and MTS (Microsoft Transaction Server), "which represent higher-level application services" [1]. Office tools, like Word and Excel, use OLE in the creation of compound documents. Web developers can automate their HTML pages using the methods, events, and properties exposed by ActiveX controls. MTS is a powerful component-based transaction processing system for building, deploying, and administering robust and secure Internet and Intranet server applications [4].

To perform the process, COM defines two principal features:

- COM uses globally unique identifiers - *class identifiers* (CLSIDs) - to uniquely identify each COM object. CLSIDs are 128-bit integers that are guaranteed to be "unique in the world across space and time" [2].

- COM objects interact with each other and with the system through collections of functions called *interfaces.* A client application has access to the object services only through the object's set of interfaces (see Figure 1). These interfaces adhere to a binary structure, which provides the basis for "interoperability between software components written in arbitrary languages" [1].

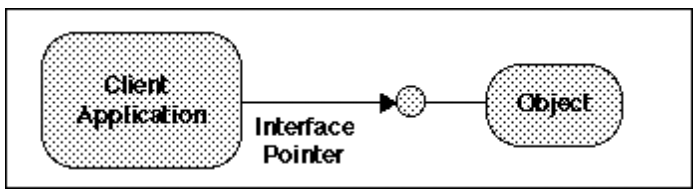

**Figure 1: Client using COM object through an Interface Pointer**

A COM interface is a strongly-typed *contract* between software components that provides a "small but useful set of semantically related operations" (methods) [3]. A COM object can support any number of interfaces. For example, Figure 2 visualizes a COM object that emulates a clock. IClock, IAlarm and ITimer are the interfaces of the clock object.

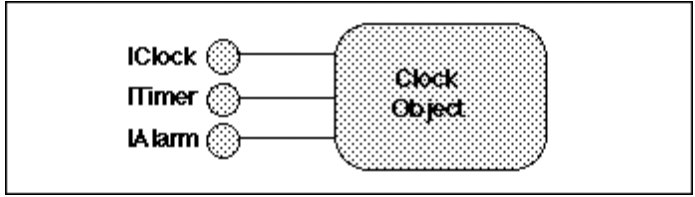

**Figure 2: Clock COM object interfaces**

Also, to perform the process, COM requires every COM object to run inside of a server. A single server can support multiple COM objects. COM objects are "either implemented within executables (EXEs) or within Dynamic Link Libraries (DLLs)" [2]. COM objects implemented in EXEs are called out-of-process servers, and these implemented in DLLs are called in-process servers [2].

The COM object life-cycle process begins with a client request to create and use a COM object. The process involves the following five steps:

1. Client request.

2. Server location.

3. Object creation.

4. Interaction.

5. Disconnection.

## 2. DESCRIPTION OF STEPS

### 2.1 Client request

The COM client request is the first phase of the object creation. Here, the COM client is every application that invokes the COM API to instantiate a new COM object. It passes a CLSID to COM, and asks for an instantiated object in return [1]. The COM client is responsible for two specific tasks, which should be included in the beginning of this phase, if they did not accomplish in the application startup:

- The COM client must verify that the COM Library version is new enough to support the functionality expected by the application. In general, an application can use an updated version of the library, but not an older one.

- The COM client must initialize the COM Library.

Regardless of the type of server in use (in-process, local, or remote), the COM client always asks COM to instantiate objects in exactly the same manner. There are two methods, which the client uses to make a request (see Figure 3). The simplest method is to call the COM function *CoCreateInstance*. This creates one object of the given CLSID, and returns an interface pointer of any requested type. Alternately, by calling *CoGetClassObject,* the client can obtain an interface pointer to what is called the *class factory* object for a CLSID. This class factory supports an interface called *IclassFactory* through which the client asks the factory to manufacture an object of its class [4].

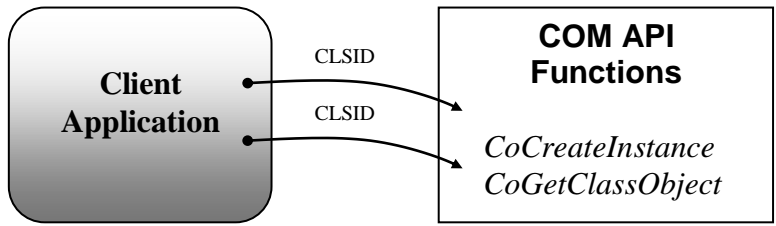

**Figure 3: The COM client request.**

## 2.2 Server location

The Server location is the next step involved in the COM object life-cycle process. In this step, "COM locates the object implementation and initiates a server process for the object" [1]. A special component called Service Control Manager (SCM) is responsible for the location and execution of the COM server that implements the COM object. The SCM ensures that when a client request is made, the appropriate server is connected and ready to receive the request. The SCM stores all the class information in the system registry, under a special text key named with the object's CLSID. For example, SCM stores the file pathnames of the COM servers, which helps for their localization. The COM client obtains this information through the COM library. This is the basis for the Server location step as shown in Figure 4.

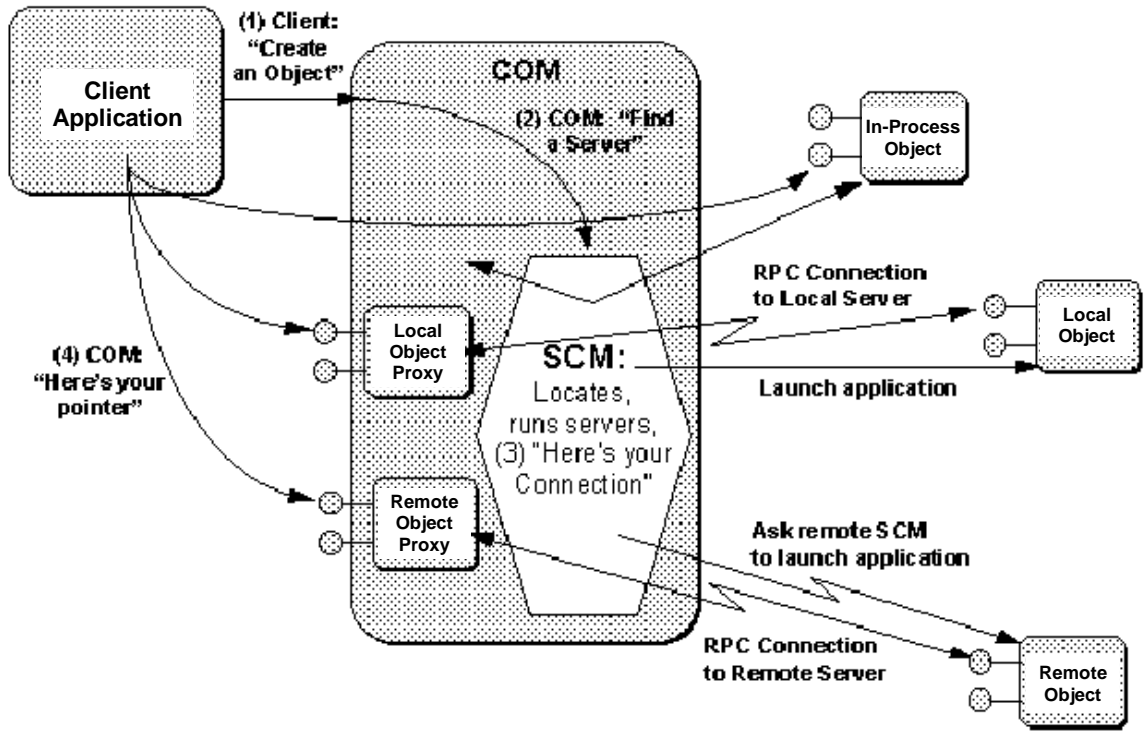

**Figure 4: COM delegates the responsibility of loading and launching servers to the SCM.**

The actions, taken by the SCM, depend on the type of COM server:

- *In-Process* - The SCM returns the file path of the DLL containing the object server implementation. The COM library then loads the DLL, and asks it for its class factory interface pointer.

- *Local* - The SCM finds and starts the local EXE, which registers a class factory interface pointer.

- *Remote* - The local SCM contacts the SCM running on the appropriate remote computer, and forwards the request to the remote SCM. The remote SCM obtains a class factory interface pointer in one of the two ways described above (in-process or local). The remote SCM then maintains a connection to that class factory, and returns an RPC connection to the local SCM [5].

The Server location is a simple task for the COM client. The complexity of this sophisticated process phase is hidden behind the SCM. The COM client does not need to perform any additional work to establish communication with a local or remote object.

## 2.3 Object Creation

The third process phase called Object Creation creates the object by given CLSID. It involves three internal steps (see Figure 5):

- Obtain the class factory for the CLSID.

- Ask the class factory to instantiate an object of the class, and return an interface pointer to the COM client.

- Initialize the COM object.

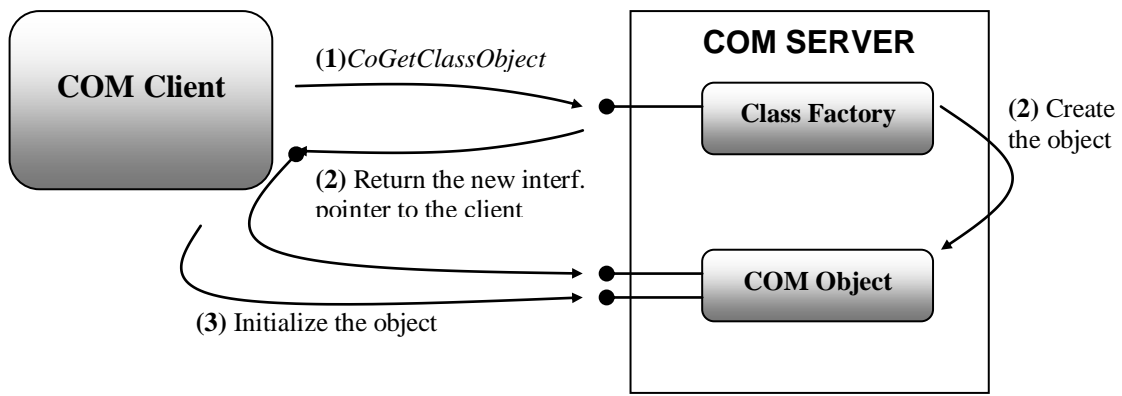

**Figure 5: The three internal steps of the Object Creation phase.**

When the COM client generates a call to the COM API function *CoGetClassObject (*the first phase of the process), this function does whatever is necessary to obtain a class factory object for the given CLSID, and returns the *IClassFactory* interface pointer to the client. After that, the client may call *IClassFactory::CreateInstance* to instantiate objects of the class. After the client has successfully created an object of a given class, it must initialize that object [5]. By definition, any newly created COM object using *IClassFactory::CreateInstance* is not initialized. Initialization generally happens through a single call to a member initialization function [4].

## 2.4 Interaction

Once an object is initialized, begins the fourth phase of the COM object life-cycle process, which is called Interaction. During this phase, the client can interact with the newly instantiated COM object through the interface pointers [1]. An interface pointer is a pointer to a function table. The Creation phase gives to the client a single interface pointer that has a limited scope of functionality. If the COM client wants to perform an operation outside that scope, it must call an interface function - *QueryInterface* - to ask for another interface on the same object. For example, the client has created and initialized an object. Now, it wants to obtain a graphical visualization from that object by calling *IMyObject::Show*. The client must call *QueryInterface* to obtain an *ImyObject* pointer before calling the function (see Figure 6). It is important to note that all the operations on that object will occur through calls to the member functions of the object's various interfaces. Every call that the client makes to *QueryInterface* to obtain another interface pointer, will internally generate a call to an interface function called *AddRef*.

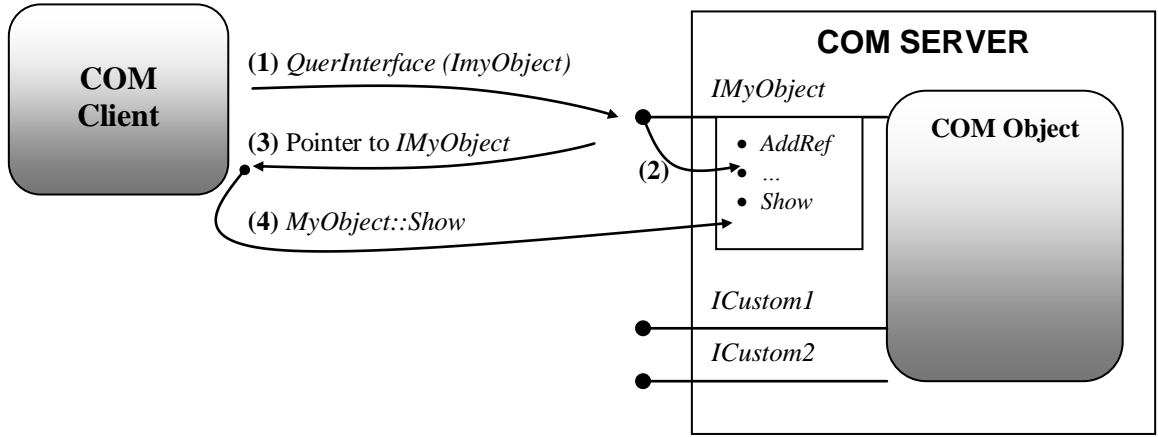

**Figure 6: The Interaction phase.**

## 2.5 Disconnection

Disconnection is the final process phase that occurs when the COM client no longer needs the COM object. The basis of this step is the *Reference Counting* mechanism. This mechanism gives to the COM object the capability to control its own lifetime. Hence, instead of freeing the object directly, the COM client must tell the object to free itself [4]. However, "the difficulty lies in having the object know when it is safe to free itself" [5]. COM specifies the R*eference Counting* mechanism to provide this control. Each object maintains a 32-bit reference count that tracks how many clients are connected to it. The use of a 32-bit counter (more than four billion clients) means that there is virtually no chance of overloading the count. Two interface functions - *AddRef* and *Release* - control the count. These two functions belong to the base COM interface *IUnknown*. *AddRef* increments the count, and *Release* decrements it. When the reference count is decremented to zero, it means that all the clients are disconnected, and the COM object *can* destroy itself (see Figure 7).

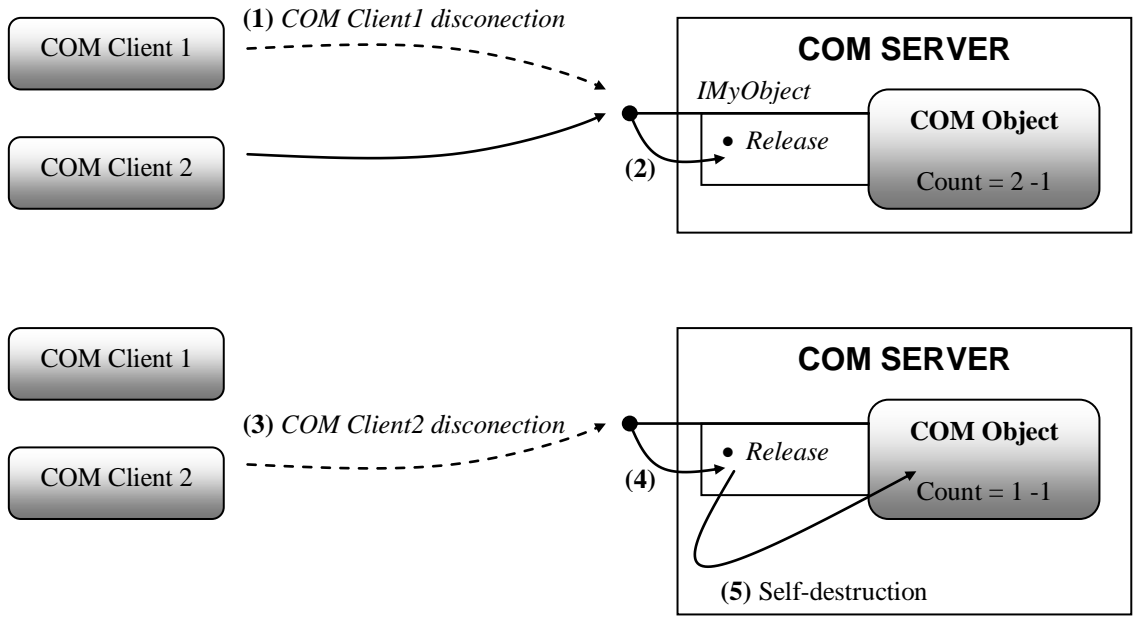

**Figure 7: The Disconnection phase for two COM clients.**

## 3. CONCLUSION

The COM object life-cycle process is the simplest process performed by COM. This process involves five main phases. The first phase begins with the COM client request for a COM object creation. When COM received the request, a special component called Service Control Manager (SCM) locates and runs the COM server that implements the COM object. The COM server performs the third phase called Object Creation. It creates the object by given CLSID. When the object is created, the COM client initializes it by calling an object's initialization function. Once an object is initialized, the client can interact with the newly instantiated COM object through its interface pointers. The client uses a special interface function – *QueryIntrface -* for obtaining these pointers. When the COM client no longer needs the COM object, the process proceeds with the last phase called Disconnection. After the client disconnection, COM uses a special mechanism - *Reference Counting -* that is responsible for the COM object lifetime.

COM is the backbone of many advanced IT technology, and its understanding gives us the advantage to be powerful IT specialists.

Process Description Of COM Object Life Cycle – Emil Vassev## REFERENCES